# Nonclassical optical response of particle plasmons with quantum informed local optics

*Weixiang Ye[#, ]**

[#] Center for Theoretical Physics, School of Physics and Optoelectronic Engineering, Hainan University, Haikou 570228, China

*Corresponding author: wxy@hainanu.edu.cn

## Abstract

As the dimensions of plasmonic structures or the field confinement length approach the mean free path of electrons, mesoscopic optical response effects, including nonlocality, electron density spill-in or spill-out, and Landau damping, are expected to become observable. In this work, we present a quantum-informed local analogue model (QILAM) that maps these nonclassical optical responses onto a local dielectric film. The primary advantage of this model lies in its compatibility with the highly efficient boundary element method (BEM), which includes retardation effects and eliminates the need to incorporate wavevector-dependent permittivity. Furthermore, our approach offers a unified framework that connects two important semiclassical theories: the generalized nonlocal optical response (GNOR) theory and the Feibelman d-parameters formalism. We envision that QILAM could evolve into a multiscale electrodynamic tool for exploring nonclassical optical responses in diverse plasmonic structures in future. This could be achieved by directly translating mesoscopic effects into observable phenomena, such as plasmon resonance energy shifts and linewidth broadening in the scattering spectrum.

# Introduction

Plasmonics, a cornerstone of nanophotonics, explores the interaction between photons and collective electron oscillations [1-3], which has evolved into an interdisciplinary research field encompassing optics, materials science, chemistry, biology, and energy over the past three decades [4-10]. This evolution has been facilitated by the adaptable nature of plasmon resonances in metal nanoparticles, known as particle plasmons [11,12]. Previous studies investigating particle plasmons in metal nanoparticles primarily relied on classical electrodynamics within the framework of the local-response approximation (LRA)[13,14]. However, as plasmonic nanostructures approach dimensions close to the mean free path of electrons, the LRA proves inadequate in delineating the optical response of metal nanostructures. Experimental observations have revealed deviations from the classical LRA description, prompting the nanophotonics community to explore the quantum origins of these phenomena [15-20]. Nonetheless, the computational demands of quantum-based theoretical frameworks, such as density-functional theory (DFT), limit their applicability to very small system sizes [21, 22]. In response, researchers have adopted semi-classical approaches that integrate quantum-informed corrections to overcome the limitations of LRA model.

One important semi-classical approach, the generalized nonlocal optical response (GNOR) theory, proposed by Mortensen et al., incorporates a classical diffusion constant ($\mathcal{D}$) in the hydrodynamic (HDM) model, allowing us to overcome our limited understanding of microscopic details related to many-body interactions[23-25]. The GNOR theory has been successfully applied to explain induced-charge screening and surface-enabled Landau damping. However, it relies on hard-wall boundary conditions and does not account for electron density spill-out or spill-in effects. Moreover, determining the value of $\mathcal{D}$ is not always straightforward. Another important approach is able to overcome the limitations imposed by hard-wall constraints is to introduce surface-response functions at the metal boundary as proposed by Peter Feibelman [26]. Feibelman d-parameters, $d_\perp(\omega)$ and $d_\parallel(\omega)$, represent the surface-response parameters that allow for the inclusion of nonlocality and electron density spill-out or spill-in effects [27,28]. Specifically, the spatial distribution of the induced surface charge can be described by $d_\perp(\omega)$, while surface conduction by surface states is accounted by $d_\parallel(\omega)$. These surface terms can be equivalently incorporated as a set of mesoscopic boundary conditions for the conventional macroscopic Maxwell equations [29,30]. A major challenge in implementing the Feibelman d-parameters theory is obtaining the Feibelman parameters for relevant metal surfaces. While progress has been made in determining Feibelman parameters for certain metal surfaces using quasi-normal-mode perturbation theory and atomic layer potential-random-phase approximation (ALP-RPA) theory [31,32], determining the Feibelman d-parameters for nanoparticles, particularly those with surface chemistry functionalization, remains a challenging task.

In this work, we present a quantum-informed local analogue model (QILAM) that offers a unified framework connecting the GNOR theory and Feibelman d-parameters theory. As the GNOR diffusion term and Feibelman d-parameters are significant only at the metal surface, a metal domain can be divided into its interior bulk part, well described by bulk optical parameters, and a thinner transition region between the bulk part and the exterior environment, which accounts for the quantum and nonlocal effects at the metal-local environment surface [29]. The dielectric permittivity of this transition region can be determined either from the electron diffusion constant and electron convection constant ($\mathcal{D}$ and $\beta$), or from the Feibelman parameters ($d_\perp(\omega)$ and $d_\parallel(\omega)$). To achieve this aim, we first find analytical expressions for the d-parameters by employing the semi-classical infinite-barrier (SCIB) model in conjunction with the GNOR theory [24,32,33]. Subsequently, we obtained the effective response functions for the metal surface covered with layers within Feibelman's formalism. Our model incorporates the influence of the local environment and electron diffusion effects with a dielectric layer, ensuring causality, compared to other local analogue models. We demonstrate that our QILAM they can be easily implemented with the boundary element method (BEM) simulation platform, as it does not require implementing the wavevector-dependent permittivity [34]. Furthermore, we compare our model with the LRA and our previous generalized nonlocal optical response theory-based local analogue model (GNORLAM) for the plasmonic optical properties of single nanorod and nanosphere dimer structures [35]. Both the QILAM and GNORLAM theories predict size-dependent linewidth ($\Gamma$) broadening and plasmon resonance energy ($E_{res}$) blue shifts in single metallic nanoparticles, as well as separation-dependent broadening of $\Gamma$ and blue shifts in $E_{res}$ for nanosphere dimer structures. However, the GNORLAM exhibits more damping effects compared to the QILAM, which is consistent with previous research [36]. The observed variation may stem from the hard wall boundary condition imposed by GNORLAM, indicating a stringent confinement of electrons within the metallic structure. Our research provides a method to connect the phenomenological GNOR parameter ($\mathcal{D}$ and $\beta$) to the frequency-dependent microscopic surface-response function in Feibelman parameters ($d_\perp(\omega)$ and $d_\parallel(\omega)$). In future, we envision that QILAM may pave the way for establishing a direct procedure a direct procedure to extract $d_\perp(\omega)$ and $d_\parallel(\omega)$ in various plasmonics structures by comparing the experimentally obtained spectral characteristics, such as $E_{res}$ and $\Gamma$ from standard dark-field spectroscopic measurements.

# Results

**Theoretical derivation of quantum-informed local analogue model.** To clarify the advancements in this contribution within the field of nanoplasmonic simulations, we delineate the connection between Hydrodynamic (HDM) theory and generalized nonlocal optical response (GNOR) theory with Feibelman d-parameters formalism, extending it further to local simulation platforms. The theoretical modeling of plasmonic phenomena in nanooptics primarily relies on Maxwell's equations, which describe the optical response of metals through constitutive relations connecting the material's response to the applied field. Mathematically, these relations are expressed as [37]:

$$\boldsymbol{D}(\boldsymbol{r},\omega) = \varepsilon_0 \int d\acute{\boldsymbol{r}}\, \varepsilon(\boldsymbol{r},\acute{\boldsymbol{r}},\omega)\boldsymbol{E}(\acute{\boldsymbol{r}},\omega) \qquad (1)$$

$$\boldsymbol{J}(\boldsymbol{r},\omega) = \int d\acute{\boldsymbol{r}}\, \sigma(\boldsymbol{r},\acute{\boldsymbol{r}},\omega)\boldsymbol{E}(\acute{\boldsymbol{r}},\omega) \qquad (2)$$

Here $\varepsilon(\boldsymbol{r},\acute{\boldsymbol{r}},\omega)$ and $\sigma(\boldsymbol{r},\acute{\boldsymbol{r}},\omega)$ represent the frequency-dependent nonlocal dielectric function and conductivity, respectively. These equations indicate that the response of a metal at a given point $\boldsymbol{r}$ depends on the electric field at neighboring points $\acute{\boldsymbol{r}}$ through the nonlocal permittivity and conductivity. In a homogeneous medium, the nonlocal permittivity depends spatially on $\boldsymbol{r} - \acute{\boldsymbol{r}}$, allowing transformation of Eq.1 and Eq.2 into $\boldsymbol{k}$-space as:

$$\boldsymbol{D}(\boldsymbol{k},\omega) = \varepsilon_0 \varepsilon(\boldsymbol{k},\omega)\boldsymbol{E}(\boldsymbol{k},\omega) \qquad (3)$$

$$\boldsymbol{J}(\boldsymbol{k},\omega) = \sigma(\boldsymbol{k},\omega)\boldsymbol{E}(\boldsymbol{k},\omega) \qquad (4)$$

The nonlocal response is clearly linked to the $\boldsymbol{k}$-dependent dielectric function and conductivity. In the case of isotropic response, the dependence solely lies on the magnitude of $\boldsymbol{k}$, regardless of its direction. Thus, to incorporate the nonlocal effect, determining how to describe the $\boldsymbol{k}$-dependent dielectric function and conductivity is imperative. A widely-used and effective method for describing the $\boldsymbol{k}$-dependent dielectric function is through the utilization of hydrodynamic (HDM) theory [25]. The nonlocal hydrodynamic treatment of electrons in metals traces its origins to seminal work by Bloch [38], with subsequent extensive literature. followed by extensive literature. This section focuses on the critical steps of the derivation and elucidates the fundamental physics of the hydrodynamic model. The Bloch hydrodynamic Hamiltonian for the inhomogeneous electron gas, denoted as $H[n(\boldsymbol{r},t), \boldsymbol{p}(\boldsymbol{r},t)]$, is defined as follows [39-41]:

$$H[n(\boldsymbol{r},t), \boldsymbol{p}(\boldsymbol{r},t)] = G[n(\boldsymbol{r},t)] + \int \frac{[\boldsymbol{p}(\boldsymbol{r},t) - e\boldsymbol{A}(\boldsymbol{r},t)]^2}{2m} n(\boldsymbol{r},t) d\boldsymbol{r}$$
$$+ e \int [\phi(\boldsymbol{r},t) + V_{back}(\boldsymbol{r})]\, n(\boldsymbol{r},t) d\boldsymbol{r} \qquad (5)$$

Here, $n(\boldsymbol{r},t)$ represents the electron density, and $\boldsymbol{p}(\boldsymbol{r},t)$ represents the conjugate momentum of the electron. The electrons interact with the electromagnetic field through the retarded potentials $\phi(\boldsymbol{r},t)$ and $\boldsymbol{A}(\boldsymbol{r},t)$. The electrostatic potential $V_{back}(\boldsymbol{r})$ acts as a confining background potential, stemmed from the electrostatic field generated by the positive ions in the metal. It can be expressed as as $\nabla^2 V_{back}(r) = \rho_{back}(r)/\varepsilon_0$, where $\rho_{back}(r)$ denotes the positive-charge density of the metal ions. Depending on the level of detail in the quantum description of the electron gas, the energy functional term $G[n(r,t)]$ includes not only the internal kinetic energy but also correlation and exchange phenomena. In the electronic fluid described by the above Hamiltonian functional, the time evolution of the state variables can be determined using Poisson brackets: $\partial_t p(r,t) = \{p(r,t), H\}$ and $\partial_t n(r,t) = \{n(r,t), H\}$. The equation governing the time evolution of the velocity field $v(r,t)$, and electron density, $n(r,t)$ is given as:

$$[\partial_t + v(r,t) \cdot \nabla]v(r,t) = -\frac{e}{m}[E(r,t) + v(r,t) \times B(r,t)] - \frac{1}{m}\nabla \frac{\delta G[n(r,t)]}{\delta n(r,t)} \quad (6)$$

In the case of plasmonic nanostructures, a simplistic approach involves considering solely the Thomas–Fermi functional and the von Weizsäcker functional, resulting in [42,43]:

$$G[n(r,t)] = \frac{3\hbar^2}{10m}(3\pi^2)^{2/3}\int n(r,t)^{5/3} dr + \frac{\hbar^2}{72m}\int \frac{\nabla n(r,t) \cdot \nabla n(r,t)}{n(r,t)} dr \quad (7)$$

Consequently, the functional derivative of $G[n]$ can be expressed as:

$$\frac{\delta G[n(r,t)]}{\delta n(r,t)} = \frac{\hbar^2}{2m}(3\pi^2)^{2/3}n(r,t)^{2/3} + \frac{\hbar^2}{36m}[\frac{1}{2}\frac{|\nabla n(r,t)|^2}{n(r,t)^2} - \frac{\nabla^2 n(r,t)}{n(r,t)}] \quad (8)$$

By combining Eq. 6 and Eq. 8 and neglecting higher-order terms of $\nabla n$ given by the von Weizsäcker functional, the convection dynamics of the electron gas is derived as:

$$[\partial_t + v(r,t) \cdot \nabla]v(r,t) = -\frac{e}{m}[E(r,t) + v(r,t) \times B(r,t)] - \gamma v(r,t) - \frac{\beta^2}{n(r,t)}\nabla n(r,t) \quad (9)$$

Here, the pressure-like term, proportional to $\beta^2 \propto v_F^2$, with $v_F$ denoting the Fermi velocity, describes a force that acts to homogenize any inhomogeneity in the electron density. This pressure term introduces nonlocal response in the hydrodynamic model. Additionally, the equation introduces a phenomenological damping term $\gamma v$, on the right-hand side for accounting for electron scattering by impurities, lattice vibrations, and other sources of electron-electron interactions.

By combining Eq.6 with the continuity equation $\partial_t n(r,t) = -\nabla \cdot [n(r,t)v(r,t)]$ and employing perturbation theory, one can derive the coupled electromagnetic equations within hydrodynamic model in the frequency domain as follows [25]:

$$\nabla \times \nabla \times E(r,\omega) = (\omega/c)^2 \varepsilon_{core}(\omega) E(r,\omega) + i\omega\mu_0 J(r,\omega) \quad (10)$$

$$[\beta^2/\omega(\omega + i\gamma)]\nabla[\nabla \cdot J(r,\omega)] + J(r,\omega) = \sigma(\omega) E(r,\omega) \quad (11)$$

Here, $J = -en_0 v$ represents the current density. In the limit of local response approximation, where $\beta \to 0$, Eq. 11 converges to the classical Ohm's law. Combining Eq. 10 and Eq.11, the governing equation in the HDM can be rewritten as:

$$\nabla \times \nabla \times E(r,\omega) = (\omega/c)^2 [\varepsilon(\omega) + \xi_{\text{HDM}}^2 \nabla(\nabla \cdot)] E(r,\omega) \tag{12}$$

Here, $\xi_{\text{HDM}}^2 = \varepsilon_{core}(\omega)\beta^2/\omega(\omega + i\gamma)$ represents the nonlocal parameter in the HDM, proportional to $v_F^2/\omega^2$. This parameter signifies the approximate distance an electron would traverse due to convection during an optical cycle. The term $\varepsilon(\omega) = \varepsilon_{core}(\omega) + i\sigma(\omega)/\varepsilon_0\omega$ represents the Drude-like permittivity of the metal, with the dielectric response from the bound electrons denoted by $\varepsilon_{core}(\omega)$. Additionally, $\sigma(\omega)$ and $\gamma$ are the Drude conductivity and damping rate of the metals, respectively.

The HDM incorporates convective current arising from the pressure term in Eq.11 while neglecting currents attributed to diffusion. To overcome this limitation, the generalized nonlocal optical response (GNOR) theory extends the HDM theory framework to incorporate electron diffusion [25]. The governing equation in the GNOR theory is expressed as follows:

$$\nabla \times \nabla \times E(r,\omega) = (\omega/c)^2 [\varepsilon(\omega) + \xi_{\text{GNOR}}^2 \nabla(\nabla \cdot)] E(r,\omega) \tag{13}$$

Here, $\xi_{\text{GNOR}}^2$ represents the GNOR nonlocal parameter that accounts for finite-range nonlocal response and is defined as by $\xi_{\text{GNOR}}^2 = \varepsilon_{core}(\omega)[\beta^2 + \mathcal{D}(\gamma - i\omega)]/\omega(\omega + i\gamma)$. The constants $\mathcal{D}$ and $\beta$ denote the induced electron diffusion constant and electron convection constant, respectively. Notably, $\xi_{\text{GNOR}}^2 = \xi_{\text{HDM}}^2 + \varepsilon_{core}(\omega)\mathcal{D}(\gamma - i\omega)/\omega(\omega + i\gamma) = \xi_{\text{HDM}}^2 + \xi_{\text{diff}}^2$, indicating the potential for multiple nonlocal mechanisms acting together to contribute to an effective nonlocal parameter. In the GNOR model, electron desnity spill-out effect is disregarded, implying $\hat{\mathbf{n}} \cdot J = 0$ on metal surfaces, suggesting no electron escape from metal volumes.

Based on the HDM and GNOR wave equations, Eq.12 and Eq.13, we can further decompose the transverse and longitudinal electric field (local and nonlocal) as follows:

$$(\nabla^2 + k_T^2)\nabla \times E(r,\boldsymbol{\omega}) = 0 \tag{14}$$

$$(\nabla^2 + k_L^2)\nabla \cdot E(r,\boldsymbol{\omega}) = 0 \tag{15}$$

Here, $k_T^2 = (\omega/c)^2 \varepsilon(\omega)$ and $k_L^2 = \varepsilon(\omega)/\xi_{\text{HDM}}^2$ or $k_L^2 = \varepsilon(\omega)/\xi_{\text{GNOR}}^2$ are the wave vectors of the transverse and longitudinal electric fields.

In our previous study, we demonstrated that the generalized nonlocal optical response (GNOR) theory can be extended to any local simulation platform (e.g., a boundary element method simulation platform) by mapping nonlocal effects onto an effective local layer with a complex dielectric function [35,44]:

$$\varepsilon_t(\omega) = \frac{\left\{[\varepsilon(\omega)]^{\frac{3}{2}}\varepsilon_b(\omega)[\omega(\omega+i\gamma)]^{\frac{1}{2}}\Delta t\right\}}{\left\langle[\varepsilon(\omega)-\varepsilon_b(\omega)]\cdot\{\varepsilon_{core}(\omega)[\beta^2+\mathcal{D}(\gamma-i\omega)]\}^{\frac{1}{2}}\right\rangle} \tag{16}$$

The thickness of the thin layer is represented by $\Delta t$, whereas the $\varepsilon_b(\omega)$ denotes the dielectric permittivity of the background. The Generalized nonlocal optical response theory-based local analogue model (GNORLAM) allows us to estimate the values of $\mathcal{D}$ and $\beta$ in single nanoparticles by comparing simulated scattering spectra with experimentally obtained data. However, we have found that the above dielectric function violates the causality condition. Causality requires that proper response functions, and clearly $\varepsilon_t(\omega)$ diverges for large ω.

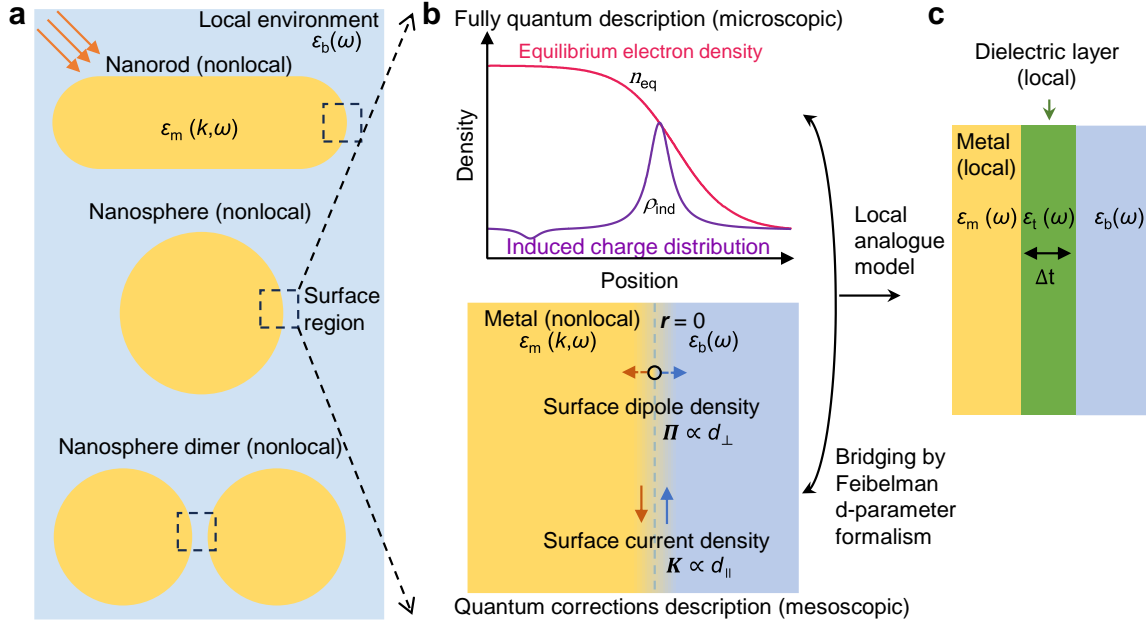

**Figure 1. Quantum-informed local analogue model (QILAM).** **(a)** Illustration depicting a light wave interacting with representative plasmonic nanostructures. The nonlocal optical response of these structures relies on a wave vector-dependent permittivity. **(b)** Furthermore, quantum and nonlocal effects have significantly impact at the interface of a metal-local environment due to the diffusive nature of the electron density profile, as illustrated in the upper panel. The lower panel demonstrates quantum surface responses characterized by the Feibelman d-parameter formalism, which rigorously incorporates quantum-mechanical effects in mesoscopic electrodynamics, bridging the gap between purely quantum (microscopic) and classical (macroscopic) domains in terms of effective current $K$ and surface polarization $\Pi$, thereby replacing the nonuniform electron density profile in the transition region. **(c)** Here, we propose a quantum-informed local model based on the Feibelman d-parameter formalism (QILAM), with a hypothetical layer (shown in green) of thickness Δt to map the quantum and non-local effects at the metal-local environment surface. The QILAM can be easily implemented with the boundary element method (BEM) simulation platform as it does not require implementing the wavevector-dependent permittivity.

To address this issue, we investigated the local analogue model within Feibelman's formalism. This approach allows us to overcome the constraints imposed by hard walls and incorporate the spill-out or spill-in effect of electron density [45]. Specifically, we examine the interaction between light wave and representative plasmonic nanostructures, such as single nanorods, nanospheres, and nanosphere dimers. The optical response of these nanostructures is nonlocal and dependent on the wavevector-dependent permittivity (**Figure 1.a**). Moreover, quantum and nonlocal effects can play a substantial role at a metal surface due to the nonuniform equilibrium electron density and the induced charge density (upper panel in **Figure 1.b**). We assume that metal surfaces appear microscopically flat when observed locally at the structure surface, denoted as *r* = 0. When optically excited, a charge distribution, represented as $p_{ind}(r)$, forms around the surface. The quantum surface responses on the metal surface then manifest macroscopically as an effective current $K$ and surface polarization $\Pi$. These responses result from $p_{ind}(r)$ and the corresponding current $J_{ind}(r)$ (lower panel in **Figure 1.c**). Within this framework, the Feibelman d-parameters, specifically $d_\perp \equiv d_\perp(\omega)$ and $d_\parallel \equiv d_\parallel(\omega)$, represent dynamical surface-response functions associated with the first moment of induced charge density and the normal derivative of tangential current density, respectively. Both parameters have an implicit ω-dependence. They can be expressed as [26,27,30]:

$$d_\perp = \frac{\int_{-\infty}^{+\infty} dr\, r\, p_{ind}(r)}{\int_{-\infty}^{+\infty} dr\, p_{ind}(r)} \tag{17}$$

$$d_\parallel = \frac{\int_{-\infty}^{+\infty} dr\, r\, \frac{dJ_{\parallel,ind}(r)}{dr}}{\int_{-\infty}^{+\infty} dr\, \frac{dJ_{\parallel,ind}(r)}{dr}} \tag{18}$$

It is important to note that the d-parameters $d_a (a = \perp/\parallel)$ are generally complex-valued. Based on this, the effective surface current and polarization can be approximated as [29,30]:

$$K = i\omega d_\parallel [\![D_\parallel]\!] \tag{19}$$

$$\Pi = d_\perp \varepsilon_0 [\![E_\perp]\!] \hat{n} \tag{20}$$

The notation $[\![f]\!] \equiv f(0+) - f(0-)$ represents the discontinuity of a field across an interface with an outward normal $\hat{n}$, and ⊥/∥ to denote the normal or parallel component of a vector. These surface terms can serve as a set of mesoscopic boundary conditions, without external interface currents or charges, for the conventional macroscopic Maxwell Equations:

$$[\![D_\perp]\!] = d_\parallel \nabla_\parallel \cdot [\![D_\parallel]\!] \tag{21}$$

$$[\![B_\perp]\!] = 0 \tag{22}$$

$$[\![E_\parallel]\!] = -d_\perp \nabla_\parallel \cdot [\![E_\perp]\!] \tag{23}$$

$$[\![H_\parallel]\!] = i\omega d_\parallel [\![D_\parallel]\!] \times \hat{n} \tag{24}$$

These mesoscopic boundary conditions represent a dual generalization from divergent perspectives, extending beyond the conventional macroscopic electromagnetic boundary conditions ($[\![D_\perp]\!] = [\![B_\perp]\!] = 0$ and $[\![E_\parallel]\!] = [\![H_\parallel]\!] = 0$) to which they reduce in the limit $d_\perp = d_\parallel = 0$.

The d-parameters are often computed using atomistic or ab-initio methods for metal-vacuum interfaces, but these methods are prohibitively time-intensive to tabulate for arbitrary metal-dielectric interfaces. We instead find analytical expressions for the d-parameters using the semi-classical infinite-barrier (SCIB) model and the hydrodynamic formalism for the longitudinal component of the dielectric tensor. From Eq.17 and 18, an intuitive physical interpretation of $d_\perp$ corresponds to the centroid of the induced charge density at the interface between two materials. Similarly, $d_\parallel$ represents the centroid of the normal derivative of the in-plane current. These general definitions allow for the computation of the Feibelman d-parameters using SCIB model, which also known as specular-reflection model (SRM). The SPM assumes specular reflection of conduction electrons at the interface [46]:

$$d_\perp^{\text{SRM}} = \frac{-2}{\pi} \frac{\varepsilon_m(\omega)\varepsilon_b(\omega)}{\varepsilon_m(\omega)-\varepsilon_b(\omega)} \int_0^\infty \frac{dk_L}{k_L^2} \left[\frac{1}{\varepsilon_L(k_L,\omega)} - \frac{1}{\varepsilon_m(\omega)}\right] \quad (25)$$

$$d_\parallel^{\text{SRM}} = 0 \quad (26)$$

In this context, $d_\parallel^{\text{SRM}} = 0$ due to the intrinsic charge-neutrality of the interface in the model. $\varepsilon_m(\omega)$ represents the Drude-like permittivity of the metal, $\varepsilon_b(\omega)$ denotes the dielectric permittivity of the local environment, and $\varepsilon_L(k_L,\omega)$ signifies he longitudinal permittivity of the spatially dispersive metal, which we compute using HDM.

We now express the $\varepsilon_m(\omega)$ and $\varepsilon_L(k_L,\omega)$ with Drude model and HDM model as follows:

$$\varepsilon_m(\omega) = \varepsilon_{core}(\omega) - \omega_p^2/(\omega^2 + i\gamma\omega) \quad (27)$$

$$\varepsilon_L(k_L,\omega) = \varepsilon_{core}(\omega) - \omega_p^2/(\omega^2 + i\gamma\omega - \beta^2 k_L^2) \quad (28)$$

Using these bulk dielectric response functions, an analytical expression for the d-parameters in the SRM can be derived utilizing Eq. 25:

$$d_\perp^{\text{SRM}} = i \frac{\varepsilon_m(\omega)\varepsilon_b(\omega)}{\varepsilon_m(\omega)-\varepsilon_b(\omega)} \frac{\beta}{\omega_p\sqrt{\varepsilon_{core}(\omega)}} \left[\frac{\varepsilon_{core}(\omega)}{\varepsilon_m(\omega)} - 1\right]^{3/2} \quad (29)$$

By expressing $\varepsilon_L(k_L,\omega)$ with GNOR theory, which substitutes $\beta$ with $[\beta^2 + \mathcal{D}(\gamma - i\omega)]^{1/2}$, we obtain:

$$d_\perp^{\text{SRM}} = i \frac{\varepsilon_m(\omega)\varepsilon_b(\omega)}{\varepsilon_m(\omega)-\varepsilon_b(\omega)} \frac{[\beta^2+\mathcal{D}(\gamma-i\omega)]^{1/2}}{\omega_p\sqrt{\varepsilon_{core}(\omega)}} \left[\frac{\varepsilon_{core}(\omega)}{\varepsilon_m(\omega)} - 1\right]^{3/2} \quad (30)$$

At this point, we would like to mention the "effective" surface-response function $d_{\text{eff}} \equiv d_\perp(\omega) - d_\parallel(\omega)$, which exhibits a "universal" behavior [47]. Firstly, it is independent of the position where the metal's electron density vanishes. Secondly, it is directly related to the spatial distribution of electron density smearing. Interestingly, this indicates that the smearing itself contributes to the electron density spill-out (for $\text{Re}(d_{\text{eff}}) > 0$) or spill-in ($\text{Re}(d_{\text{eff}}) < 0$) effects and surface-assisted Landau damping effect, which is related to $\text{Im}(d_{\text{eff}})$. We can relate $d_{\text{eff}}$ to longitudinal wave vector ($k_L$) in metal under the long-wavelength limit (in-plane wavenumber $k_\parallel \to 0$) by comparing the reflection coefficients of a flat metal-dielectric interface described by the d-parameter formalism with those described by the hydrodynamic formalism [48]:

$$d_{\text{eff}} = \frac{-i\varepsilon_b(\omega)}{k_L \varepsilon_{core}(\omega)} \frac{[\varepsilon_m(\omega) - \varepsilon_{core}(\omega)]}{[\varepsilon_m(\omega) - \varepsilon_b(\omega)]} \tag{31}$$

This expression is nearly identical to $d_{\text{eff}} = d_\perp^{\text{SRM}} - d_\parallel^{\text{SRM}} = d_\perp^{\text{SRM}}$.

For metals with a thin local surface layer of thickness $\Delta t$ and dielectric permittivity $\varepsilon_t(\omega)$, $d_{\text{eff}}$ takes the following form [48]:

$$d_{\text{eff}} = \frac{\varepsilon_b(\omega)[\varepsilon_m(\omega) - \varepsilon_t(\omega)]}{\varepsilon_m(\omega) - \varepsilon_b(\omega)} \left[\frac{1}{\varepsilon_t(\omega)} - \frac{1}{\varepsilon_b(\omega)}\right] \Delta t \tag{32}$$

By equating Eq.32 with Eq.31 for $d_\parallel(\omega) = 0$ and neglecting the high order term of $\Delta t$, we derive an analytical expression for $\varepsilon_t(\omega)$

$$\varepsilon_t(\omega) = \varepsilon_b(\omega) + \varepsilon_m(\omega) + \frac{-i\varepsilon_b(\omega)\varepsilon_m(\omega)}{\Delta t \omega_p} \left[\frac{\beta^2 + \mathcal{D}(\gamma - i\omega)}{\varepsilon_{core}(\omega)}\right]^{1/2} \left[\frac{\varepsilon_{core}(\omega)}{\varepsilon_m(\omega)} - 1\right]^{3/2} \tag{33}$$

The dielectric function above adheres to the causality condition. For large $\omega$, $\varepsilon_t(\omega) \to \varepsilon_b(\omega) + \varepsilon_m(\omega)$, as $\varepsilon_m(\omega) \to \varepsilon_{core}(\omega)$ and $\varepsilon_b(\omega), \varepsilon_m(\omega)$ are causal. Furthermore, such quantum-informed local analogue model (QILAM) integrates surface damping, and electron density spill-out or spill-in effects into a classical diffusion constant, providing a unique approach for investigating nanoparticles in larger-scale geometries beyond the scope of traditional quantum-based theoretical models. Additionally, this method solely entails the description of the local dielectric function (**Figure 1.c**), rendering it compatible with numerical calculation methods such as the boundary element method (BEM), which combines the computational efficiency of classical electrodynamics with quantum corrections.

**Feibelman d-parameters for noble metal Au.** The above quantum surface corrected description requires only the local, frequency-dependent dielectric functions, and phenomenological constants: electron diffusion constant ($\mathcal{D}$) and electron convection constant ($\beta$) as inputs. Additionally, our Feibelman $d_\perp^{\text{SRM}}$-parameters (Eq.30) are able to account for the finite compressibility of the electron gas and its influence on transverse modes through the value of $\beta$, as well as including the electronic excitations at the surface through the value of $\mathcal{D}$. Prior to applying QILAM to the nanostructure, we aim to analytically investigate the magnitude and behavior of the Feibelman $d_\perp^{\text{SRM}}$-parameters analytically for the noble metal Au as example. We varied the local environment and electron diffusion constants while maintaining the electron convection constant unchanged according to the $\beta = \sqrt{3/5}\, v_F$, which remains relatively constant for a given noble metal [49]. Besides, we also include the dynamical core-electron screening, attributable to low-lying occupied d-bands, significantly influences the optical response of noble metals. In the our model, this effect is incorporated through a polarizable background with the dielectric function $\varepsilon_{core}(\omega)$. To ensure consistency with experimental data, we construct $\varepsilon_{core}(\omega)$ by subtracting the free-electron component from the experimentally tabulated dielectric function $\varepsilon_{\text{core}}(\omega) = \varepsilon_{\text{exp}}(\omega) + \omega_p^2/(\omega^2 + i\gamma\omega)$. To achieve this, we utilize the Drude model to fit the material parameters obtained from experiments, resulting in: $\hbar\omega_{\text{p}} = 8.94$ eV, $\gamma = 0.062$ eV for noble metal Au, which are then applied in the subsequent

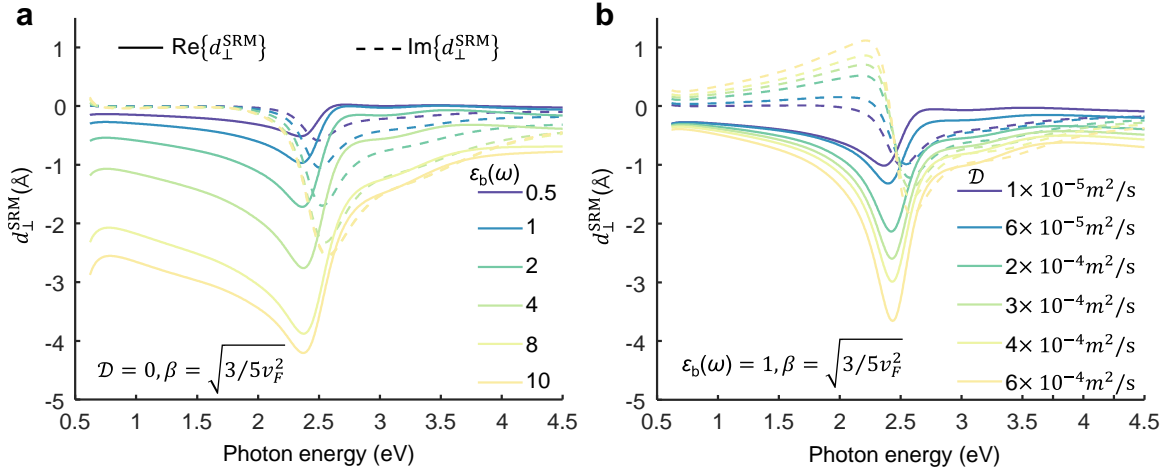

**Figure 2. Feibelman d-parameters for Au metal with different local environment and electron diffusion constants.** (a) Real (solid curves) and imaginary (dashed curves) parts of the Feibelman $d_\perp^{\text{SRM}}$-parameter computed from specular-reflection model (SRM) for various local environment with different permittivity $\varepsilon_b(\omega)$ (indicated by the color-coded legend) interfacing the metal Au with permittivity $\varepsilon_m(\omega)$ obtained experimentally. The magnitude of the real and imaginary parts of $d_\perp^{\text{SRM}}$ shifts to the lower negative value with higher values of $\varepsilon_b(\omega)$ in the absence of electron diffusion effects. Notably, all the real parts of $d_\perp^{\text{SRM}}$ are negative within the considered photon energy range, indicating electron spill-in for the conduction electrons in gold, which characterized by a relatively high work function. This phenomenon, previously associated with core electron screening in noble metals, suggests an inward shift of the centroid of the induced electron density within the metal. (b) As the values of electron diffusion effect constants ($\mathcal{D}$) increase, the magnitude of the real parts of $d_\perp^{\text{SRM}}$ shifts to the lower negative value, while the imaginary parts of $d_\perp^{\text{SRM}}$ change from negative to positive (below 2.2 eV), indicateing a more efficient decay of the plasmon into electron-hole pair excitations through surface-enabled Landau damping.

simulation [35,50].

**Figure 2.a** illustrates Feibelman $d_\perp^{\text{SRM}}$-parameters for the noble metal Au with different local environment in the absence of electron diffusion effects. The magnitude of the real and imaginary parts of $d_\perp^{\text{SRM}}$ shifts towards lower negative value with higher values of $\varepsilon_b(\omega)$ in the absence of electron diffusion effects. Notably, within the photon energy range under consideration, all the real components of $d_\perp^{\text{SRM}}$ are negative, suggesting an inward spill-in of induced electron density at the surface of the metal gold. This spill-in phenomenon in noble metals has been extensively documented in experimental studies and is attributed to screening effects from core electrons. Following, we investigated the behavior of $d_\perp^{\text{SRM}}$ with different values of $\mathcal{D}$ while fixing the dielectric functions of the local environment, and the value of $\beta$ are demonstrated in **Figure 2.b**. The magnitude of the real parts of $d_\perp^{\text{SRM}}$ shifts to the lower

negative value, while the imaginary parts of $d_\perp^{SRM}$ change from negative to positive (below 2.2 eV) with increasing values of $\mathcal{D}$. Since The imaginary part of the Feibelman parameter is related to the surface loss function and thus to the energy absorption by electronic excitations at the surface. This trend indicates a more efficient decay of the plasmon into electron–hole pair excitations through surface-enabled Landau damping, surface damping and interface damping [51-54].

**Implementation of the QILAM.** The QILAM was implemented using the boundary element method with the free software MNPBEM Toolbox to account for retardation effects. The MNPBEM Toolbox specializes in solving Maxwell's equations within dielectric environments and can model nanostructures with homogeneous and isotropic dielectric functions separated by abrupt interfaces [34]. Alternatively, other computational methods, such ascould be considered. In this simulation, tabulated dielectric function values for bulk gold were utilized. The dielectric permittivity of the thin layer was determined using Eq.33. It is important to note that the dielectric permittivity of the thin layer is not uniquely defined, as it depends on the layer thickness (Δt). We determined Δt as the maximum value of the real parts of $d_\perp^{SRM}$ in its absolute form $\Delta t = \max(|\text{Re}(d_\perp^{SRM})|)$. The reason is that $d_\perp^{SRM}$ corresponding to the centroid of the induced charge density shown in **Figure 1.b**, is situated slightly inward from the metal interface for gold. This method ensures consistency between the simulated structure and the phenomena of electron density spill-out or spill-in. Intuitively, in the electron spill-out situation (for $\text{Re}(d_\perp^{SRM}) > 0$), electrons spill out from the surface of the nanostructure, making the structure appear bigger than in the LRA model. In contrast, in the spill-in situation (for $\text{Re}(d_\perp^{SRM}) < 0$), electrons are pushed inwards into the metal, resulting in a structure that appears smaller than in the LRA. In our previous study, we determined $\beta \approx \sqrt{3/5 v_F^2} = 1.08 \times 10^6 m/s$ (with $v_F = 1.39 \times 10^6 m/s$ is the Fermi velocity in gold) and $\mathcal{D} \approx 3.0 \times 10^{-4} m^2/s$ for single gold nanorods embedded in a medium with a refractive index of $n = 1.47$. For the subsequent simulation, we used the same values, resulting in Δt =0.25 nm for $\varepsilon_t(\omega)$ in Eq.33.

**Optical response of single gold nanorods.** We apply QILAM at first to describe the optical response of single nanorods and compare it with the GNORLAM and LRA model. Two primary reasons justify choosing this nanostructure. Firstly, gold nanorods with precisely controlled sizes and aspect ratios can be readily synthesized using established protocols. Secondly, their dipole plasmon resonances can be adjusted to the near-infrared region, where inter-band damping is minimal [9-11]. We simulate the optical response of the gold nanorods (modeled as spherically capped cylinders) by irradiating them with an in-plane polarized plane wave.

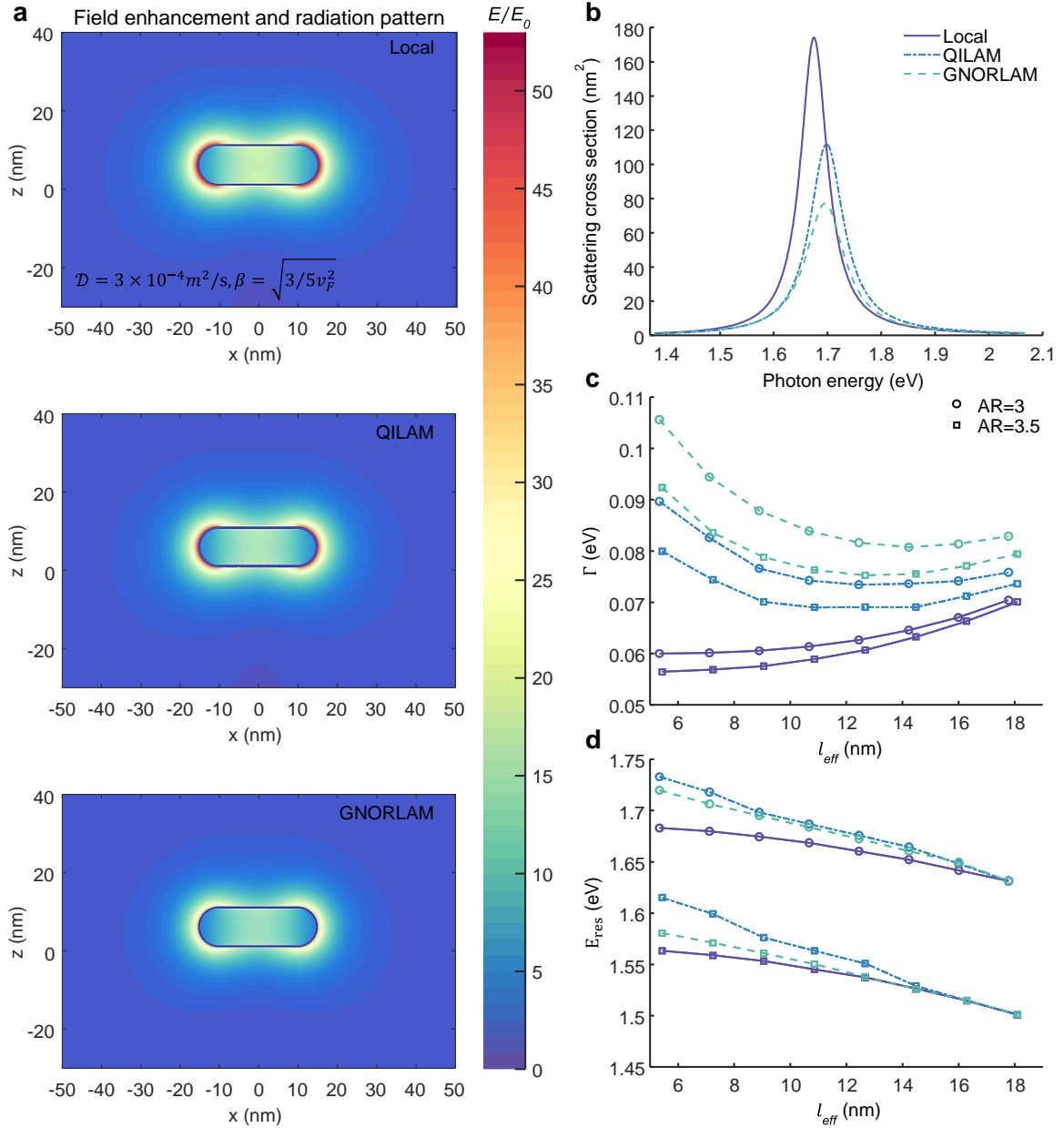

**Figure 3. Comparison of the plasmonic optical properties of gold nanorods simulated using different methods.** (a) Electric field enhancement and radiation pattern of a gold nanorod (10 nm × 30 nm) under plane wave excitation at corresponding plasmon resonance energies are depicted. The field enhancement of gold nanorods according to Quantum-informed local analogue model (QILAM) and generalized nonlocal optical response theory-based local analogue model (GNORLAM) is lower than that predicted by local-response approximation (LRA) model. All simulations use identical values for electron convection constant ($\beta$) and electron diffusion effect constant ($\mathcal{D}$). (b) Scattering spectra of the gold nanorods depicted in (a) are shown. The GNORLAM exhibits the lowest scattering cross section, consistent with its lower field enhancement. (c-d). Further simulations explored the plasmonic optical properties of gold nanorods with diameters ranging from 6 nm to 30 nm at two different aspect ratios (circle for AR = 3 and square for AR = 3.5). Plasmon resonance energy ($E_{res}$) and plasmon linewidth ($\Gamma$) were extracted from the scattering spectra and plotted against the average distance of electrons to the metal surface (or effective path length of electrons, $l_{eff}$). Both QILAM and GNORLAM predict size-dependent $\Gamma$ broadening and $E_{res}$ blue shifts in single metallic nanoparticles. However, the GNORLAM exhibits more damping compared to the QILAM. Moreover, it is notable that the simulated $\Gamma$ of smaller particles demonstrates broader behavior compared to larger particles for both QILAM and GNORLAM, suggesting a "quantum limit" to field enhancement due to quantum and nonlocal effects.

**Figure 3.a** illustrates a representative electric field enhancement and radiation pattern resulting from excitation of a gold nanorod (10 nm × 30 nm) at its respective plasmon resonance energies. The field enhancement of gold nanorods predicted by QILAM and GNORLAM is lower than that predicted by the LRA model, with GNORLAM exhibiting the lowest field enhancement. Apart from their different electric field enhancement, the corresponding scattering cross-section spectra in **Figure 3.b** also exhibit differences in peak positions, peak heights, and linewidths across the three models. The solid curve represents the LRA model, the dotted curve represents the QILAM, and the dashed curve represents the GNORLAM. Both QILAM and GNORLAM exhibit lower scattering cross-sections and blue shift effects compared to those predicted by the LRA model, consistent with the field enhancement. An intuitive explanation for this behavior can be provided based on the spread of induced charge density associated with the dipole resonance over a finite region in the near vicinity of the metal surface [25,46]. In QILAM, the induced charge spills in from the surface of the nanorods with a diffusive profile, while in GNORLAM, the induced charge is pushed inward into the metal. In contrast, in the LRA, all induced charge resides in a delta-function distribution on the interface. Consequently, plasmon polarization excitation is degraded in QILAM and GNORLAM compared with the LRA. Moreover, the blue shift of the plasmon resonance energy is linked to electron density spill-in within the metal, or more precisely, to the fact that the centroid of the induced charge density is located inside the metal. Additionally, the linewidth or damping of the plasmon resonance in QILAM and GNORLAM relates to the relaxation mechanism for the surface-induced charge. In the context of GNORLAM, the diffusion constant $\mathcal{D}$ represents the diffusive temporal spread of an initially pure surface charge into the metal volume of a plasmonic nanoparticle, a process that degrades plasmonic excitations and encompasses both mutual interactions among electrons and scattering on rough metal surfaces [25]. Moreover, it mimics surface-enhanced Landau damping resulting from the creation of electron-hole pairs. An important advantage of our semi-classical QILAM is its capability to conduct numerical computations for relatively large nanostructures. Subsequent simulations were performed to explore the optical characteristics of gold nanorods with diameters ranging from 6 nm to 30 nm and two aspect ratios (3 and 3.5). The plasmon resonance energy ($E_{res}$) and plasmon linewidth ($\Gamma$) were extracted from the simulated scattering cross-section spectrum. Without loss of generality, $\Gamma$ and $E_{res}$ of gold nanorods of gold nanorods were plotted against the average distance of electrons to the metal surface, represented as the effective path length of the electrons ($l_{eff}$) in the nanostructure. In the case of gold nanorods, the effective path length of electrons can be determined by calculating the ratio of particle volume ($V$) to surface area ($S$),

expressed as $l_{eff} = 4V/S$ [35]. As shown in **Figure 3.c** and **Figure 3.d**, demonstrate that both QILAM and GNORLAM theories predict size-dependent broadening of Γ and blue shifts of $E_{res}$ in single metallic nanoparticles. However, the GNORLAM theory exhibits a greater blue shift effect and damping compared to the QILAM theory. This difference may be attributed to the hard wall boundary condition of GNORLAM, which implies strict confinement of electrons within the metallic structure. This condition results in a uniform equilibrium density that overestimates the centroid of the induced charge density inside the metal. Interestingly, the simulated Γ in LRA model demonstrates an increasing trend with increasing particle size, whereas both QILAM and GNORLAM theories show broader behavior for smaller particles compared to larger ones, suggesting a "quantum limit" to field enhancement due to quantum and nonlocal effects. Furthermore, the blue shift effect in $E_{res}$ exhibits a decreasing trend with increasing particle size, which imply that LRA model is adequate to describe the plasmon resonance energy for large particle.

**Optical response of gold nanosphere dimers.** In our following test of the QILAM, we applied it to examine the optical response of gold nanosphere dimers and compared it with the GNORLAM and LRA models. Such nanostructures with sub-nanometer gaps are made possible by modern techniques in nanotechnology and nanofabrication, spanning from bottom-up chemically synthesized metal nanoparticles to top-down metal patterning using techniques such as electron-beam lithography (EBL) and focused-ion beam (FIB) lithography [55,56]. Similarly to the previous simulation, we modeled the optical response of the gold nanosphere dimers by irradiating them with an in-plane polarized plane wave.

**Figure 4. a** illustrates the electric field distribution and radiation pattern of gold nanosphere dimers with a diameter of 20 nm and a separation length of 2 nm under plane wave excitation at the corresponding plasmon resonance energies. The field enhancement in the interparticle gap is slightly lower in the QILAM compared to the LRA model, whereas the GNORLAM model shows the lowest field enhancement. **Figure 4.b** shows the corresponding scattering cross-section spectra across the three models. The scattering cross-section in QILAM is slightly lower than in the LRA model, and there is a transverse plasmon resonance around 2.3 eV for both models. Furthermore, the GNORLAM model exhibits the smallest scattering cross-section and the widest linewidth, consistent with its lowest field enhancement. In **Figure 4.c** and **Figure 4.d**, we compare the QILAM, GNORLAM, and LRA models for different separation distances of gold nanosphere dimers. The separation lengths range from 0.5 nm to 2 nm for two different diameters (circle for D = 10 nm and square for D = 20 nm).

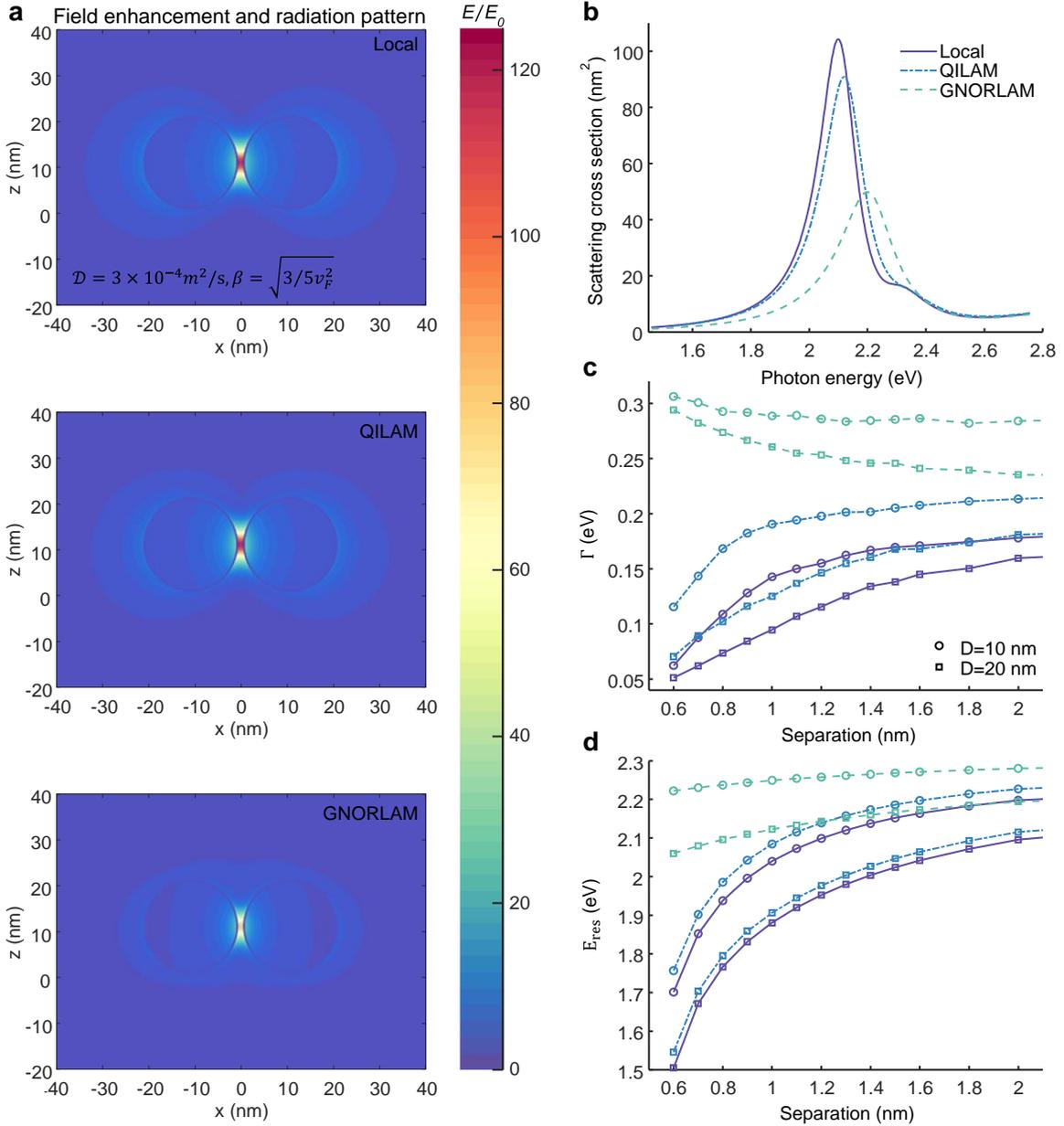

**Figure 4. Comparison of the plasmonic optical properties of gold nanosphere dimers simulated using different methods. (a)** Representation of the electric field distribution and radiation pattern of gold nanosphere dimers with a diameter of 20 nm and a separation length of 2 nm under plane wave excitation at corresponding plasmon resonance energies. The field enhancement in the interparticle gap is marginally lower for QILAM compared to the LRA model, whereas the GNORLAM demonstrates the lowest field enhancement. **(b)** The presented scattering spectra of the gold nanosphere dimers from (a) show that the GNORLAM model has the smallest scattering cross-section and the widest linewidth, in line with its lowest field enhancement. **(c-d)**. Additional simulations investigated how plasmonic optical properties depend on the separation distance of gold nanosphere dimers. The separation lengths range from 0.5 nm to 2 nm for two different diameters (circle for D = 10 nm and square for D = 20 nm). Plasmon resonance energy ($E_{res}$) and plasmon linewidth ($\Gamma$) were extracted from the scattering spectra and plotted against separation length. Both the QILAM and GNORLAM predict separation-dependent broadening of $\Gamma$ and blue shifts in $E_{res}$ for nanosphere dimer structures, indicating a "quantum limit" to field confinement at the dimer gap. Interestingly, the GNORLAM exhibits different separation-dependent trends in $\Gamma$, which the broadening of $\Gamma$ for small separation lengths. Accordingly, the GNORLAM shows greater blue shift effects compared to the QILAM.

Both the QILAM and GNORLAM theories predict separation-dependent $\Gamma$ broadening and $E_{res}$

blue shifts in nanosphere dimer structures, suggesting a "quantum limit" to field confinement at the gap of the dimer structures. Interestingly, the GNORLAM theory exhibits different separation-dependent trends in $\Gamma$, which overestimate the $\Gamma$ broadening for small separation lengths. Correspondingly, the GNORLAM theory shows greater blueshift effects compared to the QILAM theory. Again, an intuitive explanation for this behavior can be provided based on the spread of induced charge density in the near vicinity of the metal surface. In other words, the spread of induced charge density suggests that the nonlocal effect here can be interpreted as adding a separation between the two nanoparticles. However, compared with QILAM, GNORLAM overestimates the centroid of the induced charge density and relaxation rate of surface-induced charge inside the metal in gold nanosphere dimers, even though both of them use the same values of $\mathcal{D}$ and $\beta$. The above analysis and results clearly demonstrate the main strength of the QILAM. In addition to overcoming the limitations imposed by hard-wall constraints, the results presented in here also establish a solid link between surface-enhanced Landau damping and electron convection-diffusion mechanisms.

## Discussion

In conclusion, we present a quantum-informed local analogue model (QILAM) capable of describing both nonlocality and the electron density spill-in effect in the optical response of plasmonic structures. This model enables the investigation of realistic devices with relatively large particle sizes and angstrom-gap features within the classical electrodynamics framework. We demonstrate that quantum and nonlocal effects can be significant at a metal surface due to the diffusive nature of the electron density profile, which can be effectively mapped onto a local dielectric film. Effectively altering the material permittivity at the structural boundaries allows for an accurate description of nonclassical optical response in arbitrary plasmonic systems across the entire frequency range. Moreover, our approach provides a unified framework connecting two important semiclassical theories: the generalized nonlocal optical response (GNOR) theory and the Feibelman d-parameters formalism. We connect the hydrodynamic electron diffusion constant and electron convection constant to the centroid of the induced charge density at the metal interface through the Feibelman d-parameter, allowing for the extraction of the input needed for the Feibelman d-parameters formalism from experimental observable phenomena, such as resonance energy shifts and linewidth broadening in the scattering spectrum. The results obtained for a single metal-dielectric interface are applicable across various geometries, facilitating the broadening of the utility of our local analogue model in diverse applications within the realm of nonclassical plasmonics.

The primary contribution of this study lies in the advancement of a versatile local analogue model and numerical method capable of accounting for retardation effects, nonlocal effects, and quantum surface-response effects (e.g., electron spill-in or spill-out). We have not strived to extract the hydrodynamic electron diffusion constant and electron convection constant with utmost precision, either from experimentally measured values or from more microscopic calculations. Indeed, there is plenty of room for achieving even better agreement with more advanced theories. Finally, it is worth noting that the local analog approach presented in this study could further be extended to incorporate more advanced treatments at the mesoscopic level of quantum effects in plasmonics, such as incorporating the dispersive surface-response formalism, using the quantum infinite barrier (QIB) approach, and taking into account quantum tunneling effects [57-61]. Detailed investigations of these aspects are deferred for future research endeavors.

## Acknowledgements

This research was supported by the National Natural Science Foundation of China (62105229), the Humboldt Research Fellowship Programme for Experienced Researchers (CHN-1218456-HFST-E), the Hainan Provincial Natural Science Foundation of China (122RC538, 124YXQN412), and the Start-up Research Foundation of Hainan University (KYQD(ZZ)21165).

## Author Contributions

W.Y. conceived, designed, and carried out the research independently.

## References

1. Maier, S. A. Plasmonics: Fundamentals and Applications (Springer, 2007).

2. Halas, N. J. Plasmonics: an emerging field fostered by Nano Letters. Nano Lett., 10, 3816–3822 (2010).

3. Pendry, J. B., Schurig, D., Smith, D. R. Controlling electromagnetic fields. Science, 312 (5781), 1780−1782 (2006).

4. Anker, J. N. et al. Biosensing with plasmonic nanosensors. Nat. Mater., 7, 442–453 (2008).

5. Polman, A., & Atwater, H. A. Photonic design principles for ultrahigh-efficiency photovoltaics. Nat. Mater., 11, 174–177 (2012).

6. Novotny, L., van Hulst, N. Antennas for light. Nat. Photonics, 5 (2), 83−90 (2011).

7. Hao, X., Yu, M., Xing, R., Wang, C., Ye, W. Parallel frequency-domain detection of molecular affinity kinetics by single nanoparticle plasmon sensors. Appl. Phys. Lett., 121(24), 243703 (2018).

8. Ye, W., Goetz, M., Celiksoy, S., Tüting, L., Ratzke, C., Prasad, J., Ricken, J., Wegner, S. V., Ahijado-Guzmán, R., Hugel, T., Soennichsen, C. Conformational dynamics of a single protein monitored for 24 h at video rate. Nano Lett., 18(10), 6633-6637 (2018).

9. Ye, W., Celiksoy, S., Jakab, A., Khmelinskaia, A., Heermann, T., Raso, A., Wegner, S. V., Rivas, G., Schwille, P., Ahijado-Guzmán, R., Soennichsen, C. Plasmonic nanosensors reveal a height dependence of MinDE protein oscillations on membrane features. J. Am. Chem. Soc., 140(51), 17901-17906 (2018).

10. Al-Zubeidi, A., McCarthy, L. A., Rafiei-Miandashti, A., Heiderscheit, T. S., Link, S. Single-particle scattering spectroscopy: fundamentals and applications. Nanophotonics, 10, 1621-1655 (2021).

11. Sönnichsen, C., Franzl, T., Wilk, T., von Plessen, G., Feldmann, J., Wilson, O., Mulvaney, P. Drastic reduction of plasmon damping in gold nanorods. Phys. Rev. Lett., 88, 077402 (2002).

12. Yang, J., Giessen, H., Lalanne, P. Simple analytical expression for the peak-frequency shifts of plasmonic resonances for sensing. Nano Lett., 15, 3439–3444 (2015).

13. Foerster, B., Joplin, A., Kaefer, K., Celiksoy, S., Link, S., Sönnichsen, C. Chemical Interface Damping Depends on Electrons Reaching the Surface. ACS Nano, 11 (3), 2886–2893 (2017).

14. Meena, S., Celiksoy, S., Schäfer, P., Henkel, A., Sönnichsen, C., Sulpizi, M. The role of halide ions in the anisotropic growth of gold nanoparticles: a microscopic, atomistic perspective. Phys. Chem. Chem. Phys., 18, 13246-13254 (2016).

15. Uskov, A. V., Protsenko, I. E., Mortensen, N. A., O'Reilly, E. P. Broadening of plasmonic resonances due to electron collisions with nanoparticle boundary: quantum-mechanical consideration. Plasmonics, 9, 185−192 (2014).

16. Toscano, G., Straubel, J., Kwiatkowski, A., Rockstuhl, C., Evers, F., Xu, H., Mortensen, N. A., Wubs, M. Non-local optical response of metallic nanowires probed by angle-resolved cathodoluminescence spectroscopy. Nat. Commun., 6, 7132 (2015).

17. Scholl, J. A., Koh, A. L., Dionne, J. A. Quantum plasmon resonances of individual metallic nanoparticles. Nature, 483 (7390), 421−U68 (2012).

18. Raza, S., Stenger, N., Kadkhodazadeh, S., Fischer, S. V., Kostesha, N., Jauho, A.-P., Burrows, A., Wubs, M., Mortensen, N. A. Blueshift of the surface plasmon resonance in silver nanoparticles: substrate effects. Nanophotonics, 2 (2), 131−138 (2013).

19. Ciracì, C., Hill, R. T., Mock, J. J., Urzhumov, Y., Fernández-Domínguez, A. I., Maier, S. A., Pendry, J. B., Chilkoti, A., Smith, D. R. Probing the Ultimate Limits of Plasmonic Enhancement. Science, 337(6098), 1072–1074 (2012).

20. Boroviks, S., Lin, Z.-H., Zenin, V. A., Ziegler, M., Dellith, A., Gonçalves, P. A. D., Wolff, C., Bozhevolnyi, S. I., Huang, J.-S., Mortensen, N. A. Extremely confined gap plasmon modes: when nonlocality matters. Nat. Commun., 13, 3105 (2022).

21. Andersen, K., Jensen, K. L., Mortensen, N. A. & Thygesen, K. S. Visualizing hybridized quantum plasmons in coupled nanowires: From classical to tunneling regime. Phys. Rev. B 87, 235433 (2013).

22. Teperik, T. V., Nordlander, P., Aizpurua, J. & Borisov, A. G. Robust subnanometric plasmon ruler by rescaling of the nonlocal optical response. Phys. Rev. Lett. 110, 263901 (2013).

23. Raza, S., Toscano, G., Jauho, A.-P., Wubs, M. & Mortensen, N. A. Unusual resonances in nanoplasmonic structures due to nonlocal response. Phys. Rev. B 84, 121412 (R) (2011).

24. Mortensen, N. A., Raza, S., Wubs, M., Søndergaard, T. & Bozhevolnyi, S. I. A generalized non-local optical response theory for plasmonic nanostructures. Nat. Commun. 5, 3809 (2014)

25. Mortensen, N. A., Mesoscopic electrodynamics at metal surfaces - From quantum-corrected hydrodynamics to microscopic surface-response formalism, Nanophotonics, 10, 2563–2616 (2021).

26. Feibelman, P. J. Surface electromagnetic fields. Prog. Surf. Sci. 12, 287–407 (1982).

27. Liebsch, A. Electronic Excitations at Metal Surfaces (Springer, 1997).

28. Liebsch, A. Dynamical screening at simple-metal surfaces. Phys. Rev. B 36, 7378–7388(1987).

29. Yan, W., Wubs, M. & Mortensen, N. A. Projected dipole model for quantum plasmonics. Phys. Rev. Lett. 115, 137403 (2015).

30. Christensen, T., Yan, W., Jauho, A.-P., Soljačić, M. & Mortensen, N. A. Quantum corrections in nanoplasmonics: shape, scale, and material. Phys. Rev. Lett. 118, 157402 (2017).

31. Yang, Y., Zhu, D., Yan, W. et al. A general theoretical and experimental framework for nanoscale electromagnetism. Nature 576, 248–252 (2019).

32. Rodríguez Echarri, A., Gonçalves, P. A. D., García de Abajo, F. J., Tserkezis, C., Mortensen, N. A., & Cox, J. D. (2021). Optical response of noble metal nanostructures: quantum surface effects in crystallographic facets. Optica, 8(5), 710–721.

33. Ford, G. W., Weber, W. H. (1984). Electromagnetic interactions of molecules with metal surfaces. Physics Reports, 113, 195.

34. Hohenester.U.,&Trügler.A, MNPBEM - A Matlab toolbox for the simulation of plasmonic nanoparticles, Comp. Phys. Commun., 183, 370 (2012).

35. Ye, W., Nonlocal optical response of particle plasmons in single gold nanorods, Nano Lett. 23(16), 7658–7664 (2023).

36. Svendsen, M. K., Wolff, C., Jauho, A.-P., Mortensen, N. A., & Tserkezis, C. Role of diffusive surface scattering in nonlocal plasmonics. J. Condens. Matter Phys., 32(39), 395702. (2020)

37. Maxwell, J. C. A dynamical theory of the electromagnetic field. Philos. Trans. R. Soc. Lond. 155, 459–512 (1865).

38. Bloch, F. Bremsvermögen von Atomen mit mehreren Elektronen. Zeitschrift für Physik 81, 363–376 (1933).

39. Morrison, P. J. Hamiltonian description of the ideal fluid. Rev. Mod. Phys. 70, 467–521 (1998).

40. Morrison, P. J. & Greene, J. M. Noncanonical Hamiltonian density formulation of hydrodynamics and ideal magnetohydrodynamics. Phys. Rev. Lett. 45, 790–794 (1980).

41. Morrison, P. J. Hamiltonian and action principle formulations of plasma physics. Physics Plasmas 12, 058102 (2005).

42. Yang, W. Gradient correction in Thomas-Fermi theory. Phys. Rev. A 34, 4575–4585 (1986).

43. Chizmeshya, A. & Zaremba, E. Second-harmonic generation at metal surfaces using an extended Thomas-Fermi-von Weizsäcker theory. Phys. Rev. B 37, 2805–2811 (1988).

44. Luo, Y., Fernandez-Dominguez, A. I., Wiener, A., Maier, S. A., & Pendry, J. B. Surface Plasmons and Nonlocality: A Simple Model. Phys. Rev. Lett., 111, 093901 (2013).

45. Kong, J., Shvonski, A. J., & Kempa, K. Nonlocal response with local optics. Physical Review B, 97(16), 165423 (2018).

46. Gonçalves., P. A. D., Plasmonics and Light–Matter Interactions in Two-Dimensional Materials and in Metal Nanostructures: Classical and Quantum Considerations(Springer Nature, Cham, 2020).

47. Gonçalves, P. A. D., Christensen, T., Rivera, N., Jauho, A.-P., Mortensen, N. A., Soljačić, M. Plasmon–emitter interactions at the nanoscale. Nat. Commun., 11, 366 (2020).

48. Kempa, K., & Gerhardts, R. R. Nonlocal effects in ellipsometry of metallic films on metals. Surf. Sci., 150(1), 157-172 (1985).

49. Wegner, G., Huynh, D.-N., Mortensen, N. A., Intravaia, F., & Busch, K. Halevi's extension of the Euler-Drude model for plasmonic systems. Phys. Rev. B, 107, 115425 (2023).

50. Rosenblatt, G., Simkhovich, B., Bartal, G., & Orenstein, M. (2020). Nonmodal Plasmonics: Controlling the Forced Optical Response of Nanostructures. Phys. Rev. X, 10, 011071(2020).

51. G. D. Mahan, Lifetime of surface plasmons, Phys. Rev. B, 97, 075405,(2018).

52. A. Varas, P. García-González, J. Feist, F. García-Vidal, and A. Rubio, Quantum plasmonics: from jellium models to ab initio calculations, Nanophotonics, 5, 3, 409−426 (2016).

53. J. Pitarke, V. Silkin, E. Chulkov, and P. Echenique, Theory of surface plasmons and surface-plasmon polaritons, Rep. Prog. Phys., 70, 1, 1 (2006).

54. C. Yannouleas and R. Broglia, Landau damping and wall dissipation in large metal clusters, Ann. Phys.217, 1, 105−141 (1992).

55. Chen, W., Zhang, S., Kang, M., Liu, W., Ou, Z., Li, Y., Zhang, Y., Guan, Z., & Xu, H. Probing the limits of plasmonic enhancement using a two-dimensional atomic crystal probe. Light Sci. Appl., 7, 56 (2018).

56. Wang, H.-L., You, E.-M., Panneerselvam, R., Ding, S.-Y., & Tian, Z.-Q. Advances of surface-enhanced Raman and IR spectroscopies: from nano/microstructures to macro-optical design. Light Sci. Appl., 10, 161 (2021).

57. Babaze, A., Neuman, T., Esteban, R., Aizpurua, J., & Borisov, A. G. Dispersive surface-response formalism to address nonlocality in extreme plasmonic field confinement. Nanophotonics, 12(16), 3277–3289 (2023).

58. Esteban, R., Borisov, A. G., Nordlander, P. & Aizpurua, J. Bridging quantum and classical plasmonics with a quantum-corrected model. Nat. Commun. 3, 825 (2012).

59. Esteban, R. et al. A classical treatment of optical tunneling in plasmonic gaps: extending the quantum corrected model to practical situations. Faraday Discuss. 178, 151–183 (2015).

60. Hohenester, U. Quantum corrected model for plasmonic nanoparticles: A boundary element method implementation. Phys. Rev. B 91, 205436(2015).

61. Krasavin, A. V.. A brief review on optical properties of planar metallic interfaces and films: from classical view to quantum description. J. Phys. Photonics, 3, 042006 (2021).